\RequirePackage{lineno}
\documentclass[aps,prl,twocolumn,superscriptaddress]{revtex4}

\usepackage{graphicx}
\usepackage{savesym}
\savesymbol{tablenum}
\usepackage{siunitx}
\usepackage{amsmath}
\usepackage{amssymb}

\usepackage[version=4]{mhchem}

\begin{document}
\title{Machine Learning interpretation of the correlation between infrared emission features of interstellar polycyclic aromatic hydrocarbons}

\author{Zhisen Meng}
\affiliation{Laboratory for Relativistic Astrophysics, Department of Physics, Guangxi University, 530004 Nanning, China}

\author{Xiaosi Zhu}
\affiliation{Laboratory for Relativistic Astrophysics, Department of Physics, Guangxi University, 530004 Nanning, China}

\author{P\'eter Kov\'acs}
\affiliation{Institute of Materials Chemistry, TU Wien, 1060 Vienna, Austria}

\author{Enwei Liang}
\affiliation{Laboratory for Relativistic Astrophysics, Department of Physics, Guangxi University, 530004 Nanning, China}

\author{Zhao Wang}
\email{zw@gxu.edu.cn}
\affiliation{Laboratory for Relativistic Astrophysics, Department of Physics, Guangxi University, 530004 Nanning, China}

\begin{abstract}
Supervised machine learning models are trained with various molecular descriptors to predict infrared emission spectra of interstellar polycyclic aromatic hydrocarbons. We demonstrate that a feature importance analysis based on the random forest algorithm can be utilized to explore the physical correlation between emission features. Astronomical correlations between infrared bands are analyzed as examples of demonstration by finding the common molecular fragments responsible for different bands, which improves the current understanding of the long-observed correlations. We propose a way to quantify the band correlation by measuring the similarity of the feature importance arrays of different bands, via which a correlation map is obtained for emissions in the out-of-plane bending region. Moreover, a comparison between the predictions using different combinations of descriptors underscores the strong prediction power of the extended-connectivity molecular fingerprint, and shows that the combinations of multiple descriptors of other types in general lead to improved predictivity.
\end{abstract}

\maketitle

\section{Introduction}

Polycyclic aromatic hydrocarbons (PAHs) are among the most widely studied compounds in astronomy \cite{Herbst2009}, chemistry \cite{Zhang2014}, biology \cite{Moorthy2015} and environmental science \cite{Ravindra2008}. PAHs are thought to exhibit emission features dominating the infrared (IR) spectra of a large variety of galactic and extragalactic sources, commonly known as the ``unidentified'' infrared emission (UIE) features \cite{Tielens2008}. The assessment of interstellar molecular composition by IR spectroscopy crucially relies on knowledge of the bands ascribed to characteristic vibrational modes of specific bonds or functional groups in the molecule \cite{Peeters2011,Hudgins1999,Allamandola1989}. Specifically, the $3.3\,\mu$m band is due to C-H stretching, the $6.2\,\mu$m band to C-C stretching, the $7.7\,\mu$m band to coupled C-C stretching and C-H in-plane bending, the $8.6\,\mu$m band to the C-H in-plane bending and the $10-15\,\mu$m band to the CH out-of-plane bending, etc. Regardless, identifying interstellar PAHs remains anything but straightforward due to the vast variety of PAH molecular structures \cite{Li2020,Hanine2020,Qi2018}. In particular, it was often observed in interstellar spectra that the IR intensities strongly correlate with each other, the underlying cause of such cross-correlations amongst the PAH bands are poorly understood. 

For instance, strong correlations were found between the bands at $3.3$ and $11.2\,\mu$m \cite{Hony2001}, and between those at $11.3$ and $12.7\,\mu$m \cite{Acke2010}, in IR spectra from the regions of H$_\text{  I  I}$, intermediate mass star-forming, planetary nebulae and Herbig AeBe stars. These emission features are known to be associated with C-H stretching or out-of-plane bending vibration of PAHs, but their specific common structural origin is unknown. It was also reported that the $6.2$, $7.7$, $7.8$ and $16.4\,\mu$m bands exhibit strong correlations in spectra from the Milky Way, Magellanic Clouds and nearby galaxies \cite{Peeters2011}. These bands were extensively ascribed to the C-C stretching or bending mode. Likewise, cross-correlations between the $6.2$, $7.8$, $16.4$ and $8.6$ $\mu$m \cite{Vermeij2002,Acke2010,Peeters2011}, the $6.2$, $7.7$, $16.4$ and $12.7$ $\mu$m \cite{Acke2010,Keller2008,Peeters2012}, and the $11.0$ and $16.4\,\mu$m \cite{Peeters2011}, and the $11.3$ and $17.0\,\mu$m bands \cite{Smith2007} were reported for spectra from different interstellar regions, while the molecular structures responsible for these correlations are unclear. More importantly, there is a lack of quantitative measure of the band correlation for analyzing these phenomena on the ground of molecular structure or energetics, whilst there is a clear interest in the identification of the underlying connection amongst the UIE bands for assessing the possible chemical composition of interstellar medium (ISM).

A promising approach to this end is the density functional theory (DFT) calculations, which are based on a solution of the Schr{\"o}dinger equation in the Kohn-Sham formulation. DFT is able to accurately describe the changes in the molecular dipole moment with a good characterization of both the dynamics of the atomic nuclei and of the electronic charge distribution. However, a brute-force application of DFT calculations is unlikely to be successful in dealing with interstellar PAH IR spectra, due to its high computational cost vs the vast possible PAH structures existing in the emitting source. More critically for the aforementioned problem on the understanding of band correlation, the complex formulations of DFT bring great challenges for finding the molecular substructure which gives rise to a specific IR band. 

{The development of machine-learning (ML) methods has opened new and reliable ways of investigating molecular structure-spectrum relationships \cite{Ghosh2019,Butler2018}. ML was used for analysing molecular IR spectra for various applications \cite{Gastegger2017,Fu2018,Marquez2018,Ye2020,Joung2021,Calvo2021,Laurens2021,Trujillo2021}. In particular, for the analysis of interstellar IR spectra, \citet{Kovacs2020} proposed a neutral network (NN) model for predicting IR spectrum of interstellar PAHs. Their results demonstrated excellent predictive skill of the NN model based on molecular fingerprints for out-of-sample inputs. \citet{McGill2021} developed a software package for the prediction of IR spectra through the use of message passing NN based on spectra from various sources. Their method is in quantitative agreement with experiment.} 

In this work, we demonstrate that the ML model can be used to interpret the correlation between IR bands by measuring the similarity between their feature importance arrays. We take several astronomically observed correlations as examples, and show the capacity of ML in finding the molecular fragment responsible for a specific emission band, and in calculating the correlation map between bands in a spectra region. We also aim at comparing the performance of different common molecular descriptors in the ML prediction of PAH IR spectra, the choice of which is crucial for the ML prediction of molecular properties. The results of comparison are discussed in terms of the physical correlation between the IR bands and the molecular features represented by different descriptors.

\section{Methods}
\subsection{Molecular descriptors for ML}

Chemistry databases commonly use the atomic coordinate (AC) for representing a molecular compound. However for ML, AC is an ineffective descriptor because it is not invariant to translation, rotation of the molecule or the permutation of atoms. This problem can be solved by converting AC to molecular invariants, such as the extended-connectivity fingerprint (ECFP) \cite{Rogers2010}, the sorted eigenvalues (SEVs) of the distance matrix (DM) \cite{ivanciuc2000jcics} or coulomb matrix (CM) \cite{Schrier2020,rupp2012prl}. We here use six types of common molecular invariants as the descriptors, including the DM eigenvalues (DMEs), the CM eigenvalues (CMEs), ECFP, the number of H adjacency classes (NHAC) \cite{Hony2001}, the zero-point vibrational energy (ZPVE) and the atomization energy (AE), as listed in Table~\ref{T1}. These six descriptors are compared with that using a combination of molecular structural information (CMSI), which comprises the topological index meant to quantify the molecular complexity (BertzCT) \cite{Bertz1981}, etc {(obtained via \citet{rdkit})}. Among the descriptors, CMEs, DMEs, ZPVE and AE crucially rely on the DFT {or force field method} computation of bond lengths or energetics, and are thus less interesting for astronomical applications than ECFP, NHAC and CMSI that only depend upon the topology of molecule. {A voting regressor and a bootstrapping dataset treatment are also used to further improve the model predictivity, as discuss at the end of the next section.}

\begin{table}[htbp]
\centering
\caption{\label{T1}List of compared representations of molecule.}
\begin{tabular}{l|c}
\hline
Abbreviation & Descriptor name\\ \hline
DMEs & Distance matrix eigenvalues \\
CMEs & Coulomb matrix eigenvalues \\
ECFP & Extended-connectivity fingerprint\\
NHAC & Number of H adjacency classes\\
ZPVE & Zero-point vibrational energy\\
AE & Atomization energy\\
CMSI & Combined molecular structural information\\
\hline
\end{tabular}
\end{table}

\subsection{Selection of important features}

\begin{figure}[htp]
\centerline{\includegraphics[width=9cm]{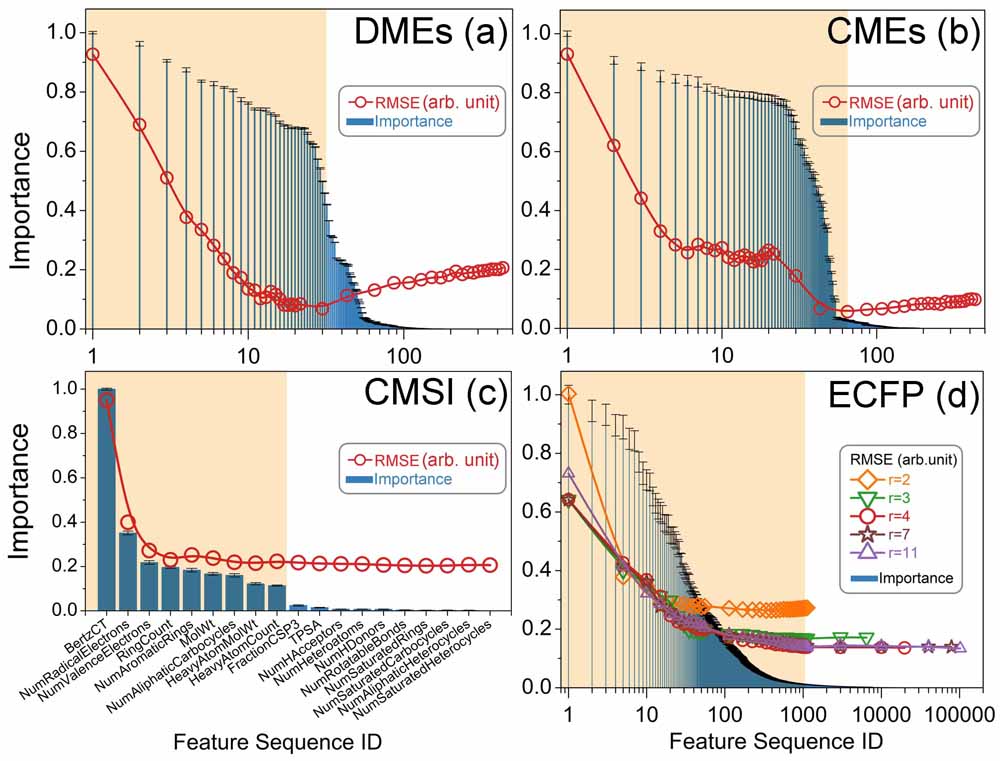}}
\caption{\label{F1}
Normalized feature importance (histogram) and root mean square error (RMSE, curves) for different descriptors. {Error bars show the mean $\pm{1}$ standard error.} The shadowed zone includes OCFs that leads to the optimal performance of the model with high efficiency. In panel (c), OCFs of CMSI include BertzCT, the number of radical electrons (NumRadicalElectrons), the number of valence electrons (NumValenceElectrons), the number of rings (RingCount), the number of aromatic rings (NumAromaticRings), the average molecular weight (MolWt), the number of aliphatic carbocycles (NumAliphaticCarbocycles), the average molecular weight ignoring hydrogens (HeavyAtomMolWt), the number of heavy atoms (HeavyAtomCount).}
\end{figure}

Using all the features available in a descriptor dataset may not result in the most effective model. A selection test is therefore performed to determine the optimal combination of features (OCFs). For each molecular descriptor, we first train a testing ML model using all features, which are subsequently sorted by their importance as shown by the histogram in Fig.~\ref{F1}. A series of models are then trained with different numbers of sorted important features (SIFs) to determine OCFs as indicated by the shadowed zone. e.g., first $30$, $65$, $9$ and $1000$ SIFs are chosen for DMEs, CMEs, CMSI and ECFP (cutoff radius $r=4$), respectively. All features are used for NHAC, ZPVE and AE which have $\leq 7$ features.

\subsection{Algorithm and dataset construction}

\begin{figure}[htp]
\centerline{\includegraphics[width=9cm]{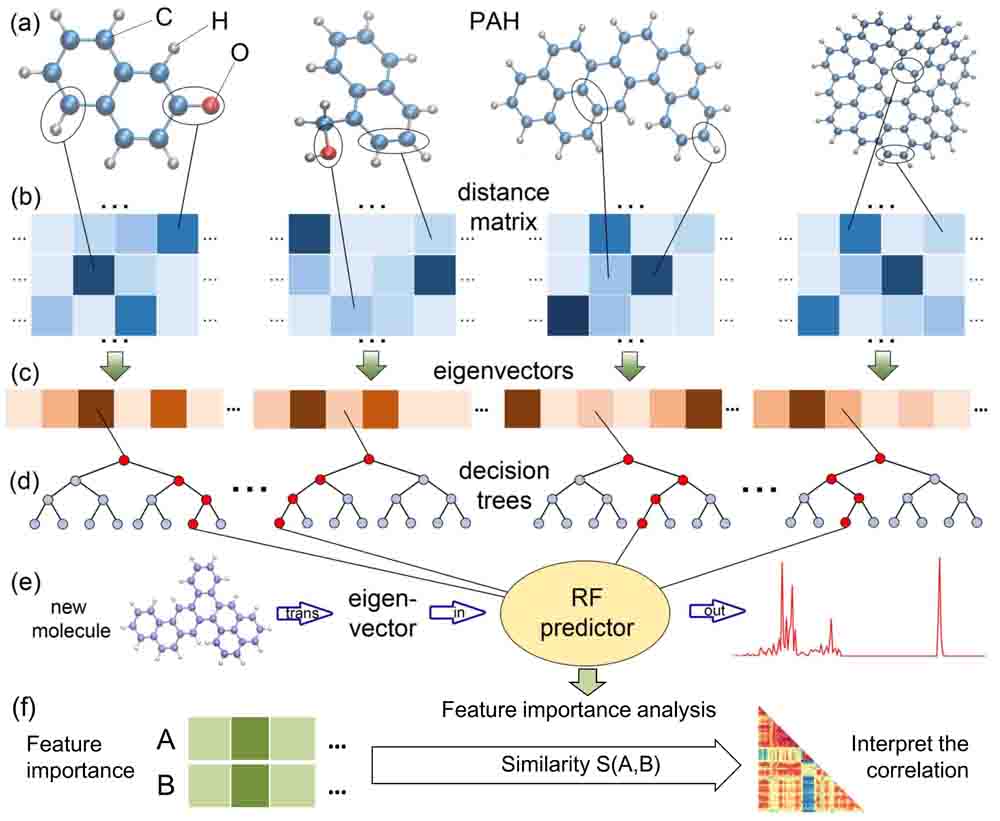}}
\caption{\label{F2}
Schematic of the ML work-flow of the DM model.}
\end{figure}

The random forest (RF) algorithm is here used to predict PAH IR spectra as implemented in the open-source Scikit-Learn library \cite{Pedregosa2011}. It is chosen because of its intrinsic metric for the importance of the feature that is very practical for the task of this study. RF is an estimator that contains a number of classifying decision trees, which comprise a root node, leaf nodes and branches \cite{Breiman2001}. When the decision trees are trained, each node will randomly select a number of specific features in which the best splitting condition is evaluated. RF uses random subsets of the features to include a number of decision networks in the learning process, which makes it efficient and naturally resistant to overfitting \cite{Ho2002}. The work-flow of our ML approach is schematically illustrated in Fig.~\ref{F2}, taking DMEs as an example. The PAH atomic coordinates are first converted to DMs (a-b), based on which SEVs are computed (c). The effective features determined by the aforementioned selection test [Fig.~\ref{F1}(a)] are used to train a RF predictor (d). The predictor is used to simulate IR spectra of the molecules in the testing set (e), and to further perform a feature importance analysis for interpreting the band correlation (f). {We use the hyperparameters adjusted in the previous work of \citet{Kovacs2020} for predicting the IR spectra of PAHs using a earlier version of the same dataset. This parameters can be found in a Python program via doi:\href{https://doi.org/10.5281/zenodo.5513073}{10.5281/zenodo.5513073}.}

$2925$ spectra are chosen from the DFT-computed spectra in the 3.20 version of the NASA AMES PAH IR spectroscopic database (PAHdb) \cite{Boersma2014nasa,Bauschlicher2018nasa,Mattioda2020nasa}. {The spectra of all charged PAHs are discarded, since we use topological descriptors such as ECFP and NHAC which could generate an exactly same input for a neutral molecule and for its charged counterpart.} The discrete IR intensity from PAHdb is converted to a histogram with a bin width of \SI{16.65}{\centi\meter^{-1}}, which is determined using Knuth's Bayesian rule as implemented in the Astropy package \cite{Knuth2006}. {Each spectrum consists of $225$ bins covering the wavelength range from 2.67 to 1172.33$\,\mu$m ($8.53$ to \SI{3746.25}{\centi\meter^{-1})}}. The histograms are separated into a low-frequency ($\upsilon_{\text{low}}$) and a high-frequency ($\upsilon_{\text{high}}$) regions by a cutoff of 4.44$\,\mu$m (\SI{2252.25}{\centi\meter^{-1}}), in order to distinguish the signals from localized vibrations of hydrogen bonds and those from other vibrations. 

A $0.8/0.2$ partition is determined by a best-predictivity test for constructing the training and the testing sets. The prediction of each model is repeated five times using different sample combinations to ensure that all samples are used in the testing and the training sets. The root mean square error (RMSE) is computed by averaging over all samples in the testing set to assess the model predictivity (namely mean RMSE). The feature importance is also evaluated to help understanding the correlation between molecular substructures and IR bands. Note that the earth mover's distance (EMD) \cite{Monge1781} is not used since this work needs a merit for the prediction of individual bands.

\section{Results and discussion}

\begin{figure}[htp]
\centerline{\includegraphics[width=9cm]{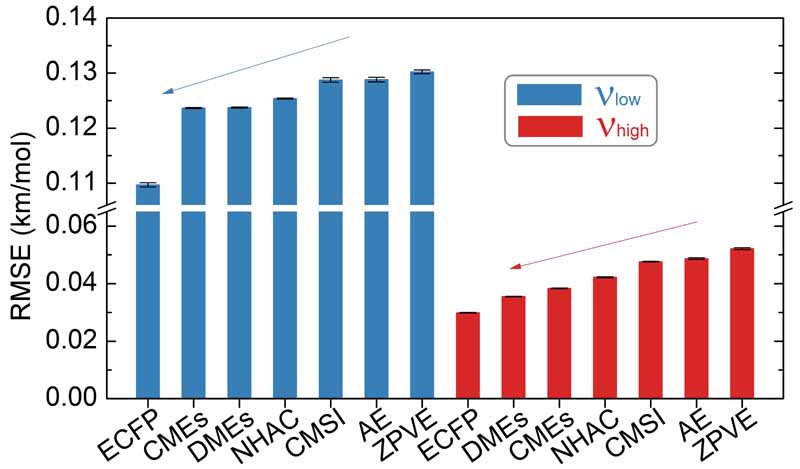}}
\caption{\label{F3}
Mean RMSE of RF prediction using different descriptors (listed in Table~\ref{T1}) in $\upsilon_{\text{low}}$ (blue bars) and $\upsilon_{\text{high}}$ (red bars). Error bars show the mean $\pm{1}$ standard error.}
\end{figure}

Fig.~\ref{F3} compares the performance of different descriptors in predicting PAH IR spectra. They are ordered by accuracy in $\upsilon_{\text{low}}$ and $\upsilon_{\text{high}}$ as follows: {ECFP \textgreater\,  CMEs \textgreater\,DMEs \textgreater\,NHAC \textgreater\,CMSI \textgreater\,AE \textgreater\,ZPVE, and ECFP \textgreater\,DMEs \textgreater\,CMEs \textgreater\,NHAC \textgreater\,CMSI \textgreater\,AE \textgreater\,ZPVE}, respectively. The prediction in $\upsilon_{\text{high}}$ is in general more accurate than that in $\upsilon_{\text{low}}$. IR bands are correlated with the vibration characteristics of specific substructures of the compound. It is hence not surprising that ECFP leads to the best predictivity since it contains the most comprehensive information about the molecular substructures amongst the compared descriptors. Besides ECFP, the atomic distance (represented by DMEs) is proven to be the most important feature in $\upsilon_{\text{high}}$, where IR emission is simply dominated by C-H stretching. It could be expected that CMEs outperform DMEs, since CMEs include not only a distance measure of the atomic configuration like DMEs does, but also additional information about the element species. This is however not the case in $\upsilon_{\text{high}}$. The species information in CMEs turns out not to be helpful, but instead likely interferes with distance information in SEV and eventually leads to a decreased predictivity.

\begin{figure}[htp]
\centerline{\includegraphics[width=9cm]{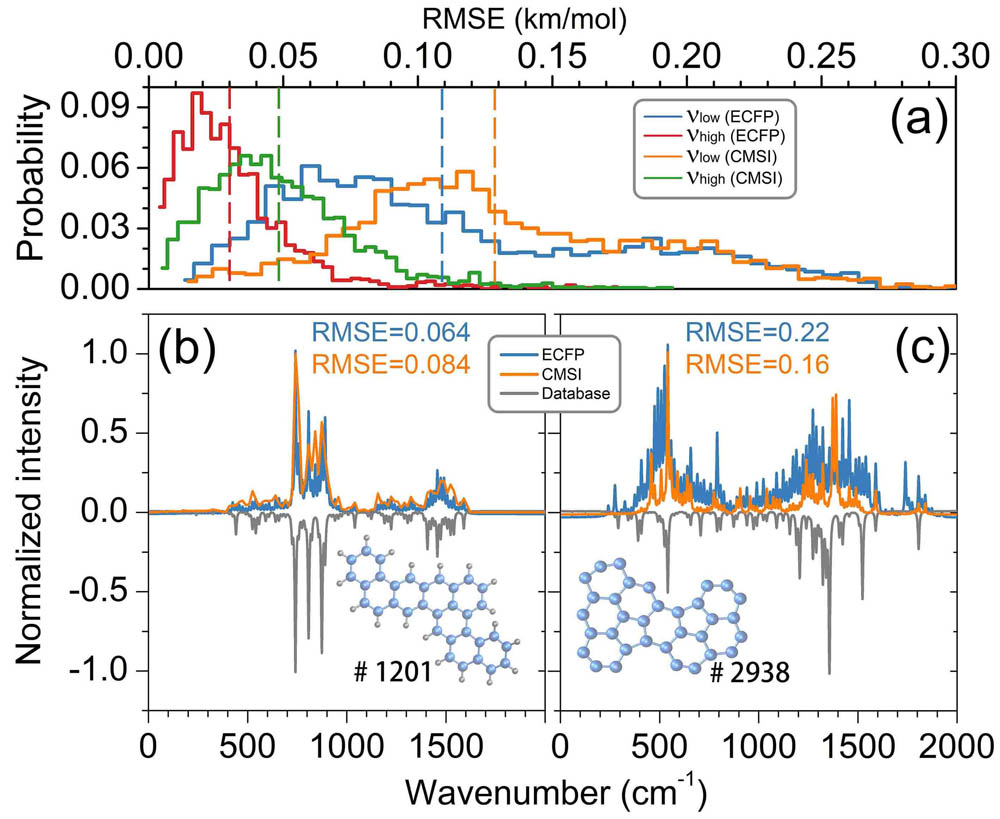}}
\caption{\label{F4}
(a) Distribution of RMSE for two models using ECFP or CMSI. The dashed vertical lines show the means. (b-c) Example IR spectra of two PAHs (\ce{C33H19} and \ce{C33}) predicted using ECFP and CMSI (colored curves, upper halves), in comparison with the referential spectra from PAHdb (gray curves, lower halves).}
\end{figure}

To give an idea about how ``good'' or ``poor'' our prediction could be, we plot in Figs.~\ref{F4} the distribution of RMSE and two examples of the predicted spectra, taking the models using ECFP and CMSI as examples. It is seen in panel (a) that RMSE is in a gamma-type distribution with a larger width for $\upsilon_{\text{low}}$. In the two example spectra, the bands of \ce{C33H19} are well predicted by ECFP in terms of both position and intensity [panel (b)]. Even the relatively poor prediction for \ce{C33} [panel (c)] is informative about the peak position. This is crucial for the subsequent analysis of the feature importance, in which the features are compared in relative to each for a specific band, despite the model may fail to match its absolute intensity in the cases of large RMSE.

\begin{figure}[htp]
\centerline{\includegraphics[width=9cm]{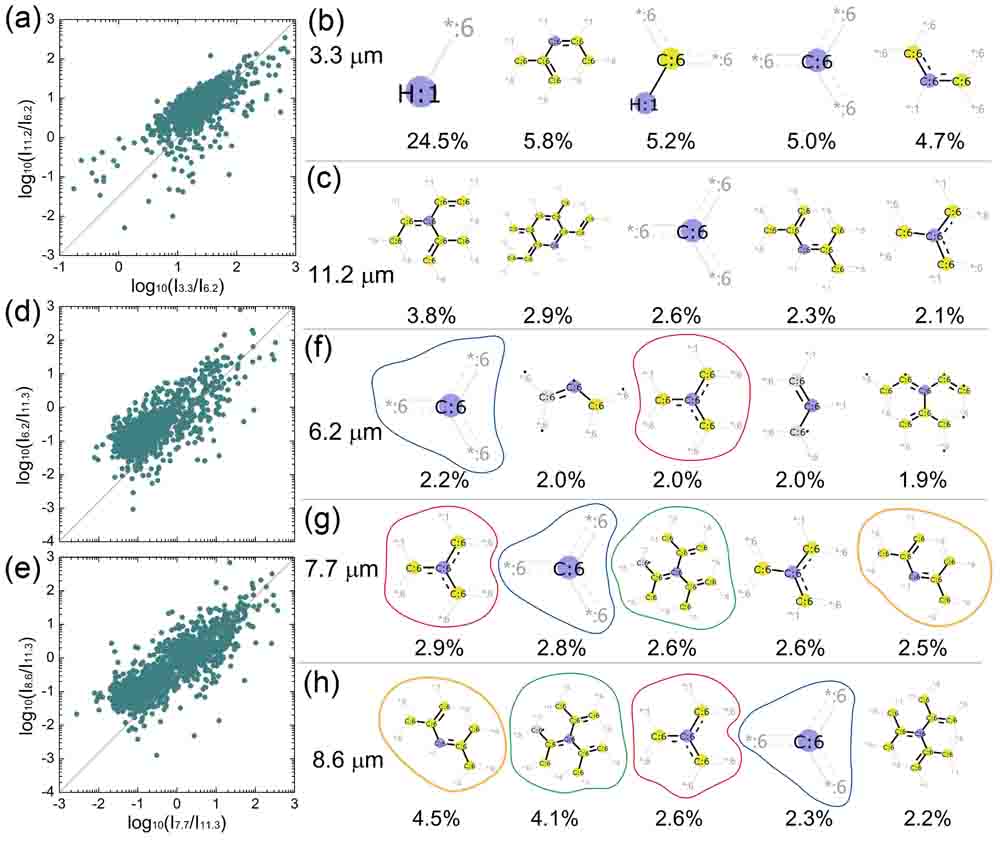}}
\caption{\label{F5}
Relative strengths of the $11.2$, $6.2$ and $8.6\,\mu$m emission features vs those at $3.3$, $7.7$ and $7.7\,\mu$m in (a), (d) and (e) respectively. The strength is normalized to that at $6.2$ or $11.3\,\mu$m. The right panels shows five major fragments of the PAHs dominating the emission features at (b) $3.3$, (c) $11.2$, (f) $6.2$, (g) $7.7$ and (h) $8.6\,\mu$m. Blue, yellow and gray circles, and asterisk represent fragment center, aromatic, nonaromatic and adjacent atoms, respectively. The digit after the colon represents the atomic number. The percentage at the bottom of each molecular fragment corresponds to the relative feature importance $I$. Coils of different colors highlight the correlated fragments.}
\end{figure}

For each band in a predicted spectrum, the importance of a feature can be computed by simply reverting all the choices based on that feature. This feature importance can be used to search for the common origins of two IR emission bands. For example, it has long been observed that the $3.3$ and $11.2\,\mu$m bands exhibit strong correlation in the spectra from H$_\text{  I  I}$ regions, intermediate mass star-forming regions and planetary nebulae \cite{Hony2001}. These bands were extensively ascribed to C-H stretching and bending modes \cite{Allamandola1989}. Our ML analysis using ECFP first confirms this correlation for the spectra collected in PAHdb [Fig.~\ref{F5} (a)], and further reveals common but slightly different structural origins of these two bands as shown in Figs.~\ref{F5} (b-c). It is seen that the $3.3\,\mu$m band is singularly dominated by the emission feature of individual C-H bonds, while the $11.2\,\mu$m one depends much more on the number of adjacent CH groups on the peripheral aromatic rings, as discussed by \cite{Hony2001}.

Another example is the strong correlation between $6.2$, $7.7$ and $8.6\,\mu$m bands previously reported by \citet{Galliano2008b} for the ISO- and Spitzer-observed mid-IR spectra from various interstellar regions. These bands were extensively ascribed to C-C stretching and C-H in-plane bending modes \cite{Allamandola1989,Draine2003}. The ML analysis confirms this correlation [Fig.~\ref{F5} (d-e)], and reveals the common molecular fragments that are responsible for these bands as those contoured by the coils in Figs.~\ref{F5} (f-h). The $8.6\,\mu$m band appears to exhibit a stronger correlation with the $7.7\,\mu$m one than the $6.2$ $\mu$m band does. However, this is not true if we compare more common features than the five ones shown in Fig.~\ref{F5} (d), given that a band can have many (up to $n=1000$ as indicated in Fig.~\ref{F1} (d)) elements in its ECFP feature importance array and each one could represent a different molecular fragment. To give a comprehensive description, we calculate the cosine similarity \cite{Singhal2001} between the feature importance arrays as an index of the band correlation,

\begin{equation}
\label{eq1}
S(A,B)=\frac{\sum\limits_{i=1}^{n}{I_{i}(A)I_{i}(B)}}{\sqrt{\sum\limits_{i=1}^{n}{I_{i}^{2}(A)}}\sqrt{\sum\limits_{i=1}^{n}{I_{i}^{2}(B)}}},
\end{equation}

\noindent where $I_{i}(A)$ is the importance of the $i$th feature of the band $A$. For the $6.2$, $7.7$ and $8.6\,\mu$m bands, $S(6.2,7.7)$, $S(7.7,8.6)$ and $S(6.2,8.6)$ are computed to be $0.79$, $0.75$ and $0.68$, respectively. In contrast to the comparison using only five features, the $6.2\,\mu$m band is actually more strongly correlated with the $7.7\,\mu$m band than the $8.6\,\mu$m band does if all $1000$ features are taken into account.

\begin{figure}[htp]
\centerline{\includegraphics[width=9cm]{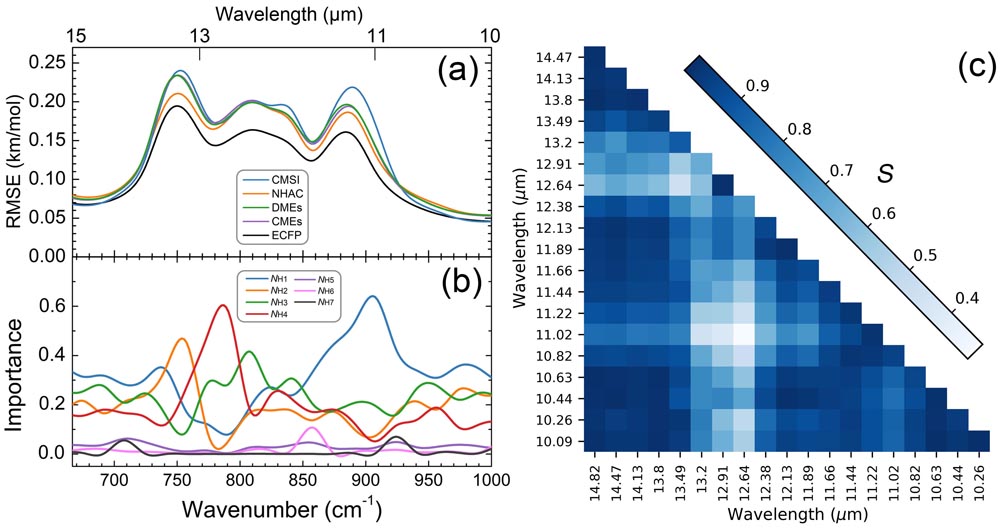}}
\caption{\label{F6}
(a) RMSE at different frequencies. (b) Frequency-dependent feature importance for the RF model using NHAC. (c) Correlation map of bands in the OOP IR region ($10-15$ $\mu$m) based on the NHAC model. Different colors correspond to different similarity (Eq.~\ref{eq1}) between the feature importance arrays of the bands.}
\end{figure}

The spectral region between $10$ and $15\,\mu$m exhibits rich emission due to the out-of-plane (OOP) bending vibrations of C-H bonds \cite{Allamandola1989}. These bands are in general difficult to predict by ML (with high RMSE), as shown in Fig.~\ref{F6}(a). The emission in this region has long been recognized to rely on the configuration of adjacent aromatic rings carrying different number of C-H bonds \cite{Bellamy1958,Hudgins1999}, which can be characterized by NHAC \cite{Hony2001}. Fig.~\ref{F5}(c) showed an example for the $11.2\,\mu$m band. Fig.~\ref{F6}(a) confirms that NHAC outperforms other descriptors except for ECFP in the OOP spectral region. However, the significance of NHAC could have been underestimated: it also gives a fair prediction for the spectrum in other IR regions.

We measure the correlation between the bands in the OOP spectral region by computing the similarity between their feature importance arrays defined by Eq.~\ref{eq1}. The result is plotted in the correlation map in Fig.~\ref{F6}(c). It is seen that the $12.64$, $12.91$, $13.20$, $11.02$ and the $11.22\,\mu$m bands are loosely correlated with the other bands, while the $14.47$ and the $10.44\,\mu$m, the $14.82$ and the $13.80\,\mu$m bands, the $14.82$ and the $10.63\,\mu$m exhibit the strongest correlations.  

\begin{figure}[htp]
\centerline{\includegraphics[width=9cm]{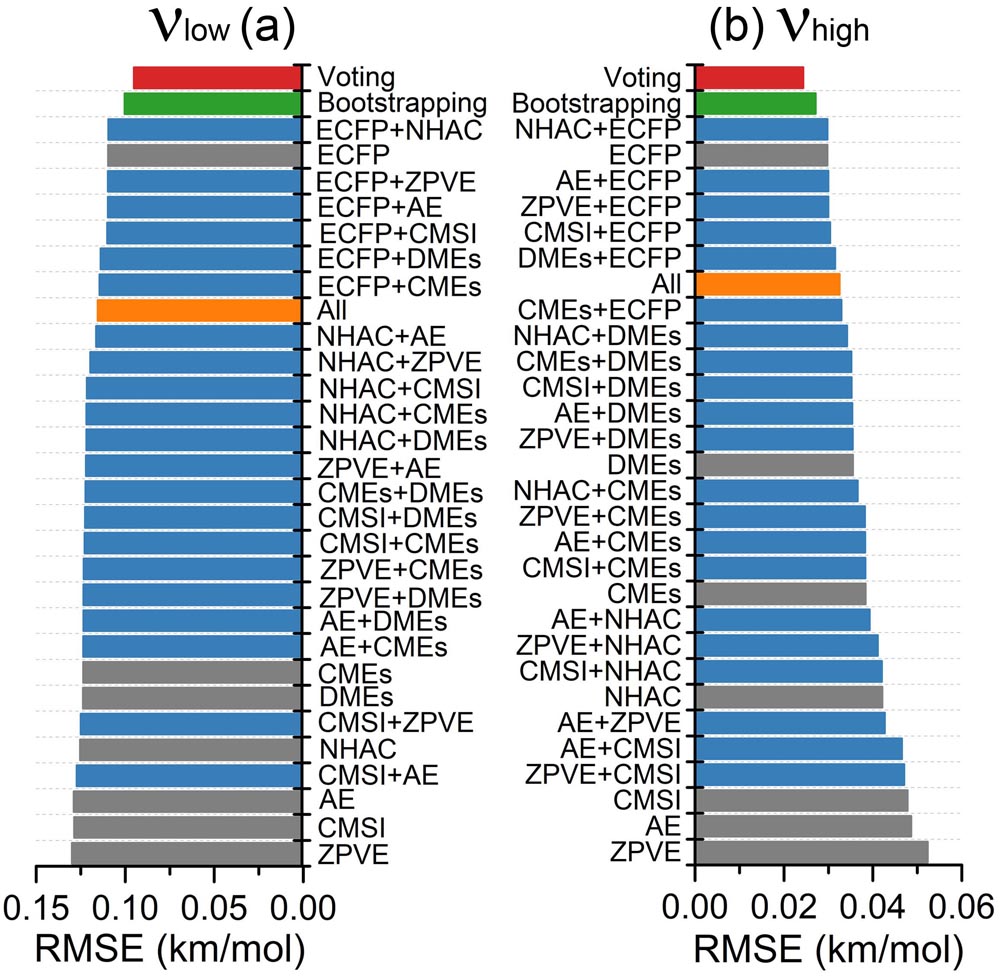}}
\caption{\label{F7}
Mean RMSE of RF prediction using different combinations of the descriptors for the PAH spectra at the low (a) and high (b) frequency regions. The descriptor combinations are sorted by their predictivity from high (top) to low (bottom). Colors are used to classify the combinations using different numbers of descriptor types.}
\end{figure}

{We perform a predictivity comparison amongst the models trained with different combinations of descriptors. These combinations are sorted by their predictivity from high (top) to low (bottom) in Fig.~\ref{F7}. It is seen that the use of the combination of different types of descriptors (blue bars) in general improves the model predictivity with respect to the model using a single type of the descriptor (gray bars). The exception of ECFP in this trend confirms the completeness of its molecular information, and underscores its importance in ML prediction of IR spectra. Besides ECFP, the combination of all descriptors (orange bars) does not lead to the best performance. Two combinations, NHAC+AE and NHAC+DMEs show good predictivity in $\upsilon_{\text{low}}$ and $\upsilon_{\text{high}}$, respectively. This indicates that the combination of two descriptors with information of complementary types could be a reliable route towards optimized predictivity. Nevertheless, a combination with AE or with DMEs is not so interesting in the goal of improving the resolution of IR spectroscopy, since these descriptor would require time-consuming calculations (e.g. via DFT). In contrast, since ECFP, NHAC and CMSI depend only on the topology of a molecule. In particular, the combination of ECFP and NHAC can be expected to be the best candidate for ML prediction of molecular IR spectra.}

{The methods of the plurality voting \cite{Lin2003} and the bootstrapping dataset \cite{Kohavi1995} are also used in order to further improve the model predictivity. The former method is a voting regressor that combines every model built with a different descriptor, it performs individual prediction with the most votes to form a final prediction. The latter method is an ensemble meta-estimator that fits base regressors each on random subsets of the original dataset, it summarizes the individual predictions in a voting fashion to form a final prediction. Fig.~\ref{F7} shows that these two methods lead to further improved predictivity, as indicated by the red and green bars. Moreover, we compute the degrees of mutual information between the different descriptor families and those between the features within the families, using the maximal information coefficient (MIC) algorithm \cite{Reshef2011}. It is found that ECFP and NHAC are loosely correlated with the other descriptors, as shown in doi:\href{https://doi.org/10.5281/zenodo.5513073}{10.5281/zenodo.5513073}. The features of ECFP and NHAC exhibit relatively weak cross-correlation within the family, indicating less repeated information. This could be parts of the reasons why ECFP outperforms other descriptors, and why the descriptor pairs with NHAC exhibit improved predictivity.}

\section {Conclusions}

Different ML models are trained to predict IR spectra of PAHs using several different types of representation of the compounds. We demonstrate that ML can help interpreting the correlation between IR emission features. The astronomically observed correlations between the $3.3$ and $11.2\,\mu$m bands, and between the $6.2$, $7.7$ and $8.6\,\mu$m ones are used as examples for demonstration. The RF feature importance is computed to find the common responsible molecular fragments, from which the correlation can be quantified by measuring the cosine similarity of their feature importance arrays (Eq.~\ref{eq1}). It is found that the $3.3\,\mu$m band is singularly dominated by the emission feature of individual C-H bonds, while the $11.2\,\mu$m one depends much more on the local atomic environment (i.e. the number of adjacent CH groups on the peripheral aromatic rings). It is also shown that the $6.2\,\mu$m band is more strongly correlated with the $7.7\,\mu$m one than the $8.6\,\mu$m band does. The correlation map of the OOP spectral region on the ground of NHAC suggests that the $12.64$, $12.91$, $13.20$, $11.02$ and the $11.22\,\mu$m bands have weak correlations, and that the $14.47$ and the $10.44\,\mu$m bands, the $14.82$ and the $13.80\,\mu$m bands, the $14.82$ and the $10.63\,\mu$m bands exhibit strong correlations.  

We compare the performance of the descriptors, the choice of which is crucial for the model predictivity. ECFP has the best performance amongst the individual descriptors in both the low and the high frequency regions, underscoring the completeness of its contained molecular information. The second place of DMEs in $\upsilon_{\text{high}}$ indicates a strong correlation between the atomic distance and the high-frequency spectra. NHAC outperforms other descriptors (except for ECFP) for the bands in the $10-15\,\mu$m region. Moreover, besides ECFP, the combination of two descriptors with information of complementary types leads to improved predictivity. {The voting regressor and the bootstrapping dataset are shown to be reliable methods to further improve the model predictivity.} Our results underscore the potential of ECFP and NHAC for accurately analysing astronomical IR spectra, since they encode the topology of the molecule without reference to the exact coordinates of the atoms or the energy of molecule that need to be obtained by specific quantum chemical calculations.

\section*{Acknowledgements}
G. K. H. Madsen and J. Carrete are acknowledged for helpful discussions. Partial financial support from the National Natural Science Foundation of China (11964002) is acknowledged.


\end{document}